\documentclass[aip,jcp,reprint,amsmath,amsfonts,amssymb,a4paper,eqsecnum,superscriptadress]{revtex4-1}
\usepackage{color,graphicx}
\newtheorem{definition}{Definition}%[section] 
\newtheorem{lemma}{Lemma}%[section] 
\newtheorem{proposition}{Proposition}
\newtheorem{example}{Example}%[section]
\newcommand{\qed}{\hfill\ensuremath{\Box}}

\begin{document}

\title{Geometry and Statistics of Ehrenfest dynamics}
\author{J. L. Alonso}
\affiliation{Departamento de F\'{\i}sica Te\'orica, Universidad de Zaragoza,  Campus San Francisco, 50009 Zaragoza (Spain)}
\affiliation{Instituto de Biocomputaci{\'{o}}n y F{\'{\i}}sica de Sistemas Complejos (BIFI), Universidad de Zaragoza,  Edificio I+D, Mariano Esquillor s/n, 50018 Zaragoza (Spain)}
\author{A. Castro}
\protect
\affiliation{Instituto de Biocomputaci{\'{o}}n y F{\'{\i}}sica de Sistemas Complejos (BIFI), Universidad de Zaragoza,  Edificio I+D, Mariano Esquillor s/n, 50018 Zaragoza (Spain)}
\author{J. Clemente-Gallardo}
\affiliation{Departamento de F\'{\i}sica Te\'orica, Universidad de Zaragoza,  Campus San Francisco, 50009 Zaragoza (Spain)}
\affiliation{Instituto de Biocomputaci{\'{o}}n y F{\'{\i}}sica de Sistemas Complejos (BIFI), Universidad de Zaragoza,  Edificio I+D, Mariano Esquillor s/n, 50018 Zaragoza (Spain)}
\author{J. C. Cuch\'{\i}}
\affiliation{Departament d'Enginyeria Agroforestal, ETSEA- Universitat de Lleida,
  Av. Alcalde Rovira Roure 191, 25198 Lleida  (Spain)}
\author{P. Echenique}
\affiliation{Instituto de Qu\'{\i}mica F\'{\i}sica ``Rocasolano'' (CSIC), C/ Serrano 119, 28006 Madrid (Spain)}
\affiliation{Departamento de F\'{\i}sica Te\'orica, Universidad de Zaragoza, Campus San Francisco, 50009 Zaragoza (Spain)}
\affiliation{Instituto de Biocomputaci{\'{o}}n y F{\'{\i}}sica de Sistemas Complejos (BIFI), Universidad de Zaragoza,  Edificio I+D, Mariano Esquillor s/n, 50018 Zaragoza (Spain)}
\author{F. Falceto}
\affiliation{Departamento de F\'{\i}sica Te\'orica, Universidad de Zaragoza,  Campus San Francisco, 50009 Zaragoza (Spain)}
\affiliation{Instituto de Biocomputaci{\'{o}}n y F{\'{\i}}sica de Sistemas Complejos (BIFI), Universidad de Zaragoza,  Edificio I+D, Mariano Esquillor s/n, 50018 Zaragoza (Spain)}

\begin{abstract}
  Quantum dynamics (e.g., the Schr{\"{o}}dinger equation) and classical
  dynamics (e.g., Hamilton equations) can both be formulated in equal
  geometric terms: a Poisson bracket defined on a manifold. The
  difference between both worlds is due to the presence of extra
  structure in the quantum case, that leads to the appearance of the
  probabilistic nature of the measurements and the indetermination
  and superposition principles. In this paper we first show that
  the quantum-classical dynamics prescribed by the Ehrenfest equations
  can also be formulated within this general framework, what has been
  used in the literature to construct propagation schemes for
  Ehrenfest dynamics.  Then, the existence of a well defined Poisson bracket
  allows to arrive to a Liouville equation for a statistical
  ensemble of Ehrenfest systems. The study of a generic toy model
  shows that the evolution produced by Ehrenfest dynamics is ergodic
  and therefore the only constants of motion are functions of the
  Hamiltonian. The emergence of the canonical ensemble characterized
  by the Boltzmann's distribution follows after an appropriate
  application of the principle of equal a priori probabilities to this
  case. This work provides the basis for extending stochastic methods
  to Ehrenfest dynamics.
\end{abstract}

\maketitle

\section{Introduction}
\label{sec:introduction}

  There is little doubt that the solution of
  Schr{\"{o}}dinger equation for a combined system of electrons and
  nuclei enables us to predict most of the chemistry and molecular
  physics that surrounds us, including bio-physical processes of great
  complexity. Yet there is also little doubt that this task is not
  possible in general, and approximations need to be made; one of the
  most important and successful being the classical approximation for
  a number of the particles. Mixed quantum-classical dynamical (MQCD)
  models are therefore necessary and widely used.

  The so-called ``Ehrenfest equations'' (EE) result from a
  straightforward application of the classical limit to a portion of
  the particles of a full quantum system, and constitute the most
  evident MQCD model, as well as a first step in the intricated
  problem of mixing quantum and classical dynamics. For example, it
  can be noted that much of the field of molecular dynamics (MD) --
  whether ab initio or not -- is based on equations that can be
  obtained from further approximations to the Ehrenfest model: for
  instance, written in the adiabatic basis, the Ehrenfest dynamics collapse into
  Born Oppenheimer MD if we assume the non-adiabatic
  couplings to be negligible.

  It is not the purpose of this work, however, to dwell into the
  justification or validity conditions of the EE (see for example
  Refs.~\onlinecite{marxhutter:2009, gerber1982time, gerber1988self,
    bornemann1996quantum, bornemann1995pre} for rigurous
  analyses). Nor is it to investigate the unsettled problem of which
  is the best manner of mixing quantum and classical degrees of
  freedom. The Ehrenfest model has a proven application niche, and for
  this reason we are interested in investigating some of its
  theoretical foundations, in a manner and with an aim that we
  describe in the following.~\footnote{For recent
      progress in non-adiabatic electronic dynamics in MQCD see, for
      example, C. Zhu, A. W. Jasper and D. G. Truhlar, J. Chem. Theor. Comp. {\bf 1}, 527
      (2005).}

Classical mechanics (CM) can be formulated in several mathematical
frameworks, each corresponding to a different level of abstraction
(Newton's equations, the Hamiltonian formalism, the Poisson brackets,
etc.). Perhaps its more abstract and general formulation is
geometrical, in terms of Poisson manifolds. Similarly, quantum
mechanics (QM) can be formulated in different ways, some of which
resemble its classical counterpart. For example, the observables
(self-adjoint linear operators) are endowed with a Poisson algebra
almost equal to the one that characterizes the dynamical variables in
CM.  Moreover, Schr{\"{o}}dinger equation can be recast into
Hamiltonian equations form~\cite{heslot1985quantum} by transforming
the complex Hilbert space into a real one of double dimension; the
observables are also transformed into dynamical functions in this new
phase space, in analogy with the classical case. Finally, a Poisson
bracket formulation has also been established for QM, which permits to
classify both the classical and the quantum dynamics under the same
heading.

  This variety of formulations does not emerge from academic caprice;
  the successive abstractions simplify further developments of the
  theory, such as the step from microscopic dynamics to statistical
  dynamics: the derivation of Liouville's equation (or von Neumann's
  equation in the quantum case), at the heart of statistical dynamics,
  is based on the properties of the Poisson algebra.

 The issue
  regarding what is the correct equilibrium distribution of a mixed
  quantum-classical system is seen as a very relevant
  one.~\cite{Par2006JCTC, Kab2006JPCA, Bas2006CPL, Kab2002PRE,
    Tul1998Book, Mul1997JCP, Mau1993EPL, Ter1991JCP} An attempt to its
  derivation can be found in Ref.~\onlinecite{Mau1993EPL}, where they
  arrive to the same distribution that we will advocate below,
  although it is done using the Nos{\'{e}}-Hoover
  technique,\cite{nose:1984,nose:1991} which is only a mathematical scheme to
  produce the equilibrium, and which the very authors of
  Ref.~\onlinecite{Mau1993EPL} agree that it is not clear how to apply
  to a mixed quantum-classical dynamics. In
  Ref.~\onlinecite{Par2006JCTC} they provide some analytical results
  about this distribution, but not in the case in which the system of
  interest is described by the EE; instead (and as in
  Refs.~\onlinecite{Kab2006JPCA, Bas2006CPL}), they assume that the
  system is fully quantum, and that it is coupled to an infinite
  classical bath via an Ehrenfest-like interaction.

  It is therefore necessary to base Ehrenfest dynamics -- or any other
  MQCD model -- on firm theoretical ground. In particular, we are
  interested in establishing a clear path to statistical mechanics for
  Ehrenfest systems, which in our opinion should be done by first
  embedding this dynamics into the same theoretical framework used in
  the pure classical or quantum cases (i.e., Poisson brackets,
  symplectic forms, etc). Then the study of their statistics will
  follow the usual steps for purely classical or purely quantum
  ensembles. In this respect, it should be noted that other approaches
  to MQCD (not based on Ehrenfest equations) exist,\cite{Kisil:fk,
    prezhdo1997mixing, kapral1999mixed} and to their corresponding
  statistics;~\cite{nielsen2001statistical,Kapral:2001fk} however it
  has been found that the formulation of well defined
  quantum-classical brackets (i.e., satisfying the Jacobi identity and
  Leibniz derivation rule) is a difficult
  issue.~\cite{kisil2005quantum,agostini2007we,kisil2010comment,agostini2010reply}
  On the contrary, as shown below, and as a result of the fact that
  the evolution of a quantum system can be formulated in terms of
  Hamilton equations similar to those of a classical system,\cite{bornemann1996quantum,Schmitt1996JCP} we will
  not have difficulties in a rigorous formulation of the Ehrenfest
  dynamics in terms of Poisson brackets.

%The scope of this work is more limited than other approaches to
%Mixed Quantum-Classical Dynamics (MQCD) \cite{Kisil:fk,prezhdo1997mixing,
%kapral1999mixed} and its corresponding 
%statistical \cite{nielsen2001statistical,Kapral:2001fk}. We are only
%interested in the dynamics and statistics of Ehrenfest. 

%Among the difficulties that appear in those MQCD
%\cite{Kisil:fk,prezhdo1997mixing,kapral1999mixed} we highlight the 
%formulation of a quantum-classical brackets satisfying the Jacobi
%identity and the Leibniz chain rule
%\cite{kisil2005quantum,agostini2007we,kisil2010comment,agostini2010reply}.  
 
%Contrary, as shown below, and as a result of the fact that the
%evolution of a quantum system can be formulated in terms of
%Hamilton's equations similar to  those of a classical system, we will
%not have difficulties in a rigorous formulation of the Ehrenfest
%dynamics in term of Poisson  brackets. Then the study of their
%statistics will follow the usual steps for systems purely classical or
%purely quantum. 

%The aim of this paper is thus twofold:
%provide a rigorous Poisson 
%description  of Ehrenfest dynamics and, by using it, build the
%corresponding statistical description. 

  The roadmap of this project is the following: In
  Section~\ref{sec:ehrenfest} we recall the definition of the
  Ehrenfest model. In Section \ref{sec:append-geom-dynam-PB} we
  summarize quickly the formulation of CM in terms of Poisson
  brackets. Then we summarize, in
  Section~\ref{sec:append-geom-quant}, the analogous description of QM
  in terms of geometrical objects which can be found, for example, in
  Refs.~\onlinecite{
      kibble1979geometrization,heslot1985quantum,abbati1984pure,
      cirelli1991quantum,brody2001geometric,Ashtekar:1998p906,
      Carinena:2006p7565,Carinena:2007p813,clemente2008basics}. These
  works demonstrate how Schr{\"{o}}dinger  equation can be written as
  a set of (apparently classical) Hamiltonian equations. Also, by
  using a suitable definition of observables as functions on the
  (real) set of physical states, Schr\"odinger  equation can be
  written in terms of a canonical Poisson bracket. QM appears in this
  way as ``classical'', although the existence of extra algebraic
  structure encodes the probabilistic interpretation of measurements,
  and the superposition and indetermination principles \cite{
      kibble1979geometrization,abbati1984pure,
      cirelli1991quantum,brody2001geometric,Ashtekar:1998p906} . 
  For the case of Ehrenfest dynamics, we can now
  combine it (Section~\ref{sec:our-proposal}) with CM, whose Poisson
  bracket formulation is very well known. The procedure is very
  simple: we consider as a global phase space the Cartesian product of
  the electronic and the nuclear phase spaces and define a global
  Poisson bracket simply as the sum of the classical and the quantum
  ones. It is completely straightforward to prove that such a Poisson
  bracket is well defined and that it provides the correct dynamical
  equations (EE).

  From a formal point of view, the resulting dynamics
  is more similar to a classical system than to a quantum one,
  although when considering a pure quantum system the dynamics is the
  usual Schr\"odinger one. This is not surprising since the coupling
  of the classical and quantum systems makes the total system
  nonlinear in its evolution, and this is one of the most remarkable
  differences between a classical and a quantum system in any
  formulation.

  Once the dynamical description as a Poisson system is
  at our disposal, it is a very simple task
  (Section~\ref{sec:defin-stat-syst}) to construct the corresponding
  statistical description following the lines of
  Refs.~\onlinecite{balescu1997statistical} and
  \onlinecite{balescu1975equilibrium}. The formal similarity with
  the CM case ensures the correctness of the procedure and allows us
  to derive a Liouville equation.  The choice of the equilibrium
  distribution is based, as usual, in the principle of equal
  \textit{a priori} probablities implicitely used by Gibbs and clearly
  formulated by Tolman \cite{tolman}.

  Finally, in the conclusions (Section
  \ref{sec:concl-future-work}), we propose to extend, within our
  formalism, stochastic methods to Ehrenfest dynamics.

\section{The Ehrenfest model}
\label{sec:ehrenfest}

The Ehrenfest Equations have the following general form:
\begin{eqnarray}
\dot{\vec{R}}_J(t) & = & \langle \psi(t) \vert \frac{\partial
  \hat{H}}{\partial \vec P_J}(\vec R(t),\vec P(t),t) \vert \psi(t)\rangle,
\\
\dot{\vec{P}}_J(t) & = & -\langle \psi(t) \vert \frac{\partial
  \hat{H}}{\partial \vec R_J}(\vec R(t),\vec P(t),t) \vert \psi(t)\rangle,
\\
i\hbar \frac{{\rm d}}{{\rm d}t} |\psi(t)\rangle & = & \hat{H}(\vec R(t), \vec
P(t),t)\vert\psi(t)\rangle,
\label{eq:ehrenfest}
\end{eqnarray}
where $(\vec R, \vec P)$ denote collectively the set of canonical position and momenta
coordinates of a set of classical particles, whereas $\psi$ is the
wavefunction of the quantum part of the system. See Ref ~\onlinecite{Andrade:2009p7108} for the issue of the Hellmann-Feynman theorem
in this context -- i.e., whether one  should take the derivative inside or outside the expectation value. 

This and other MQCD models appear in different contexts; in many
situations, the division into quantum and classical particles is made
after the electrons have been integrated out, and is used to
\emph{quantize} a few of the nuclear degrees of freedom. However, a
very obvious case to use EE is when we want to treat electrons quantum
mechanically, and nuclei classically. In order to fix ideas, let us
use this case as an example: the Hamiltonian for the full quantum
system is:
\begin{align}
  \label{eq:29a}
  \hat H=&-\hbar^2\sum_J\frac 1{2M_J}\nabla_J^2-\hbar^2\sum_j\frac 12
  \nabla_j^2+\frac 1{4\pi \epsilon_0}\sum_{J<K}\frac{Z_JZ_K}{|\vec R_J-  
  \vec R_K|}
  \notag
  \\
  &-\frac 1{4\pi \epsilon_0}\sum_{j<k}\frac 1{|\vec r_j-\vec r_k|}-\frac 1{4\pi \epsilon_0}\sum_{J,j}\frac {Z_J}{|\vec R_J-\vec r_j|} 
  \notag 
  \\
  =&-\hbar^2\sum_J\frac 1{2M_J}\nabla_J^2 -\hbar ^2\sum_j\frac 12
 \nabla_j^2+V_{n-e}(\vec r,\vec R) 
 \notag
 \\
 =&-\hbar^2\sum_J\frac 1{2M_J}\nabla_J^2+H_e(\vec r, \vec R),
\end{align}
where all sums must be understood as running over the whole natural
set for each index. $M_J$ is the mass of the J-th nucleus in units of
the electron mass, and $Z_J$ is the charge of the J-th nucleus in
units of (minus) the electron charge. Also note that we have defined
the nuclei-electrons potential $V_{n-e}(\vec r ,\vec R)$ and the
electronic Hamiltonian $H_e(\vec r, \vec R)$ operators.

The EE may then be reached in the following way:\cite{marxhutter:2009,gerber1982time,gerber1988self,bornemann1996quantum,bornemann1995pre}
first, the full wave function is split into a product of nuclear and
electronic wave functions, which leads to the time dependent
self-consistent field model, in which the two subsystems are quantum
and 
coupled. Afterwards, a classical limit procedure is applied to the
nuclear subsystem, and the EE emerge naturally.
In terms of the nuclei positions $\vec R_{J}$ and of the element of the Hilbert space 
$|\psi\rangle \in \mathcal{H}$ which encodes the state of the
electrons of the system, the Ehrenfest dynamics is then given by
\begin{align}
  \label{eq:21a}
  M_{J}\ddot{\vec R}_{J}&=-\langle\psi | \nabla_{J}H_e(\vec r, \vec R) |\psi \rangle,
 \\
  i\hbar\frac{d}{dt} |\psi\rangle&=H_e(\vec r, \vec R) |\psi\rangle.
\end{align}

These equations can be given a Hamiltonian-type description by introducing a Hamiltonian function of the form: 
\begin{equation}
  \label{eq:22a}
  H(\vec R, \vec P)=\sum_{J}\frac{\vec P_{J}^2}{2M_J}+\langle\psi | H_e(\vec r,\vec R) |\psi \rangle.
\end{equation}

Then, by fixing a relation of the form $ \vec P_{J}= M \dot {\vec
  R}_{J}$, we obtain a structure similar to Hamilton equations:
\begin{align}
  \label{eq:25a}
  \dot{\vec R}_{J}&=\frac {\vec P_{J}}{M_{J}},
  \\
  \dot {\vec P}_{J}&=-\langle\psi | \nabla_{J}H_e(\vec r, \vec R) |\psi \rangle,
  \\
 i\hbar \frac{d}{dt} |\psi\rangle&=H_e(\vec r, \vec R) |\psi\rangle.
\end{align}
But in spite of the formal similarities, these equations do not correspond
yet to Hamilton equations, since they lack a global phase space
formulation (encompassing both the nuclear and the electronic degrees
of freedom) and a Poisson bracket.

\section{Classical mechanics in terms of Poisson brackets}
\label{sec:append-geom-dynam-PB}

Let us begin by recalling very quickly the Hamiltonian formulation of
classical dynamics. We address the interested reader to a classical
text, such as Ref.~\onlinecite{Abraham:1978p1131}, for a more detailed
presentation. 

Let us consider a classical system with phase space $M_C$, which, for
the sake of simplicity, we can identify with $\mathbb{R}^{2n}$, where
$n$ is the number of degrees of freedom (strictly speaking, $M_C$ is a
general $2n$-dimensional manifold, homeomorphic to $\mathbb{R}^{2n}$
only locally). The $2n$ dimensions correspond with the $n$ position
coordinates that specify the configuration of the system, and the $n$
corresponding momenta (mathematically, however, the division into
``position'' and ``momenta'' coordinates is a consequence of Darboux theorem -- 
see below).
%In that manifold, there are two types of degrees of
%freedom: the states of the physical variables describing the position
%of the system (by ``position'' we mean any relevant degree of freedom
%one should consider), and their corresponding momenta. We shall use
%$(\vec R, \vec P)$ as notation to represent these variables.

In what regards the observables, classical mechanics uses the set of
differentiable functions
\begin{equation}
  \label{eq:37}
  f:M_C\to \mathbb{R},
\end{equation}
 assigning the result of the measurement to every point in $M_C$. 
On this set of functions $C^\infty(M_C)$ we introduce an operation,
known as Poisson bracket, which allows us to study the effect
of symmetry transformations and also the dynamical evolution.  The
precise definition is as follows:
\begin{definition}
 A \textbf{Poisson bracket}, $\{\cdot,\cdot\}$, is a bilinear operation 
\begin{equation}
  \label{eq:36}
  \{ \cdot, \cdot\}:C^\infty(M_C)\times C^\infty(M_C)\to C^\infty(M_C), 
\end{equation}
which:%satisfies
\begin{itemize}
\item It is antisymmetric, 
$$
\{f, g\}=-\{g, f\}, \qquad \forall f,g\in C^\infty(M_C).
$$
\item It satisfies the Jacobi identity, i.e., $\forall
f,g,h\in C^\infty(M_C)$:
$$
\{ f, \{g, h\}\}+ \{ h, \{f, g\}\}+\{ g, \{h, f\}\}=0.
$$
\item It satisfies the Leibniz rule ,i.e., $\forall
f,g,h\in C^\infty(M_C)$:
$$
\{f, gh\}=\{f,g\}h+g\{f,h\}.
$$
\end{itemize}
\end{definition}
 
A Poisson bracket allows us also to introduce the concept of
Hamiltonian vector field:
\begin{definition}
  Given a function $f\in C^\infty(M_C)$ and a Poisson bracket
  $\{\cdot, \cdot \}$, a vector field, $X_f$, is said
  to be its \textbf{Hamiltonian vector field} if 
$$
X_f(g)=\{f,g\}, \qquad \forall g\in C^\infty(M_C).
$$
\end{definition}
\begin{definition}
We shall call a \textbf{Hamiltonian system} to a triple
$(M_C, \{\cdot, \cdot\}, H)$, where $\{ \cdot, \cdot\} $ is a Poisson
bracket on $M_C$, and dynamics is introduced via the function $H\in
C^\infty(M_C)$, that we call the Hamiltonian. 
\end{definition}
If the Poisson bracket is non degenerate (it has no Casimir
functions), a theorem due to Darboux ensures that there exists a set
of coordinates $(\vec{R},\vec{P})$ for which the bracket has the ``standard'' form
(at least locally, in a neighbourhood of every point):
\begin{equation}
\{ f_1,f_2\}=\sum_{k=1}^n \frac{\partial f_1}{\partial P_k}\frac{\partial
    f_2}{\partial R^k}- \frac{\partial f_1}{\partial R^k}\frac{\partial
    f_2}{\partial P_k}.
\end{equation}
These coordinates are called Darboux coordinates. They are specially useful when studying
the dynamics and invariant measures of a Hamiltonian system\cite{Abraham:1978p1131}. 
In the rest of the paper we shall be working with this kind of 
coordinates.

Then, given $f\in C^\infty(\mathbb{R}^{2n})$,
  we can write the corresponding Hamiltonian vector field $X_f$ as
\begin{equation}
    X_f=\sum_{k=1}^n \frac{\partial f(R,P)}{\partial P_k}\frac{\partial}{\partial R^k}-
\frac{\partial f(R,P)}{\partial R^k}\frac{\partial}{\partial P_k}.
\end{equation} 

%\begin{example}
%  Let $M_C=\mathbb{R}^2$ with coordinates $(R, P)$. We consider as Poisson bracket
%$$
%\{ f_1,f_2\}=\frac{\partial f_1}{\partial P}\frac{\partial
%    f_2}{\partial R}- \frac{\partial f_1}{\partial R}\frac{\partial
%    f_2}{\partial P}.
%$$ 

%Then, given $f\in C^\infty(\mathbb{R}^2)$,
%  we can write the corresponding Hamiltonian vector field $X_f$ as
%  \begin{equation*}
%    X_f=\frac{\partial f(R,P)}{\partial P}\frac{\partial}{\partial R}-
%\frac{\partial f(R,P)}{\partial R}\frac{\partial}{\partial P}.
%  \end{equation*} 
%\end{example}

%The geometric formulation of Hamiltonian Mechanics is
%very often defined on Poisson manifolds, i.e. manifolds
%endowed with a Poisson bracket on the corresponding space of
%functions. 
Now one can consider two different formulations of the dynamics:
\begin{itemize}
 \item One which defines the corresponding Hamiltonian vector
       field $X_H$ obtained as above
       $$
       X_H(g)=\{ H, g\}, \qquad \forall g\in C^\infty(M_C).
       $$
       The integral curves of the vector field $X_H$ define the solution of the dynamics.
 \item An analogous formulation can be given in terms of the
       observables. If we consider now the set of functions of the
       system, i.e., the set of classical observables,
       %which contains, as elements,
       %the functions `position' and `momenta' of each particle (i.e. $\vec R$
       %and $\vec P$), 
       the dynamics is written as the Poisson bracket of the
       Hamiltonian function $H$ with any other function of the system, i.e.,
      \begin{equation}
        \label{eq:23}
        \frac{df}{dt}=\{ H, f\},\qquad \forall f\in C^\infty(M_C).
      \end{equation}
\end{itemize}
Both approaches are equivalent: the differential equations that determine
the integral curves of the Hamiltonian vector field are given by Eq.~\ref{eq:23}, 
for the functions ``position'' and ``momenta'' of each particle (i.e., $\vec R$ and $\vec P$). These equations
are nothing else but Hamilton equations:
\begin{equation}
  \label{eq:8}
  \begin{cases}
\dot R^j=\displaystyle \frac{\partial H}{\partial P_j},\\
\dot P_j=\displaystyle -\frac{\partial H}{\partial R^j}
\end{cases}
\end{equation}

%Both approaches are equivalent, as we can see in the following example:
%\begin{example}
%Consider $M_C$ and the Poisson bracket defined as 
%$$
%\{ f_1,f_2\}=\sum_k \frac{\partial f_1}{\partial P_k}\frac{\partial
%    f_2}{\partial R^k}- \frac{\partial f_1}{\partial R^k}\frac{\partial
%    f_2}{\partial P_k}.
%$$ 
%Consider also a Hamiltonian of the form
%$$
%H(\vec R, \vec P)= \sum_k\frac {P_k^2}{2M_k}+V(\vec R),
%$$
%and the corresponding vector field $X_H$:
%$$
%X_H=\sum_{j }\frac{\partial H}{\partial P_j} \frac{\partial }{\partial R^j}-
%\frac{\partial H}{\partial R^j}\frac{\partial }{\partial P_j},
%$$
%whose integral curves are the solutions of Hamilton equations:
%\begin{equation}
%  \label{eq:8}
%  \begin{cases}
%\dot R^j=\frac{\partial H}{\partial P_j}=\frac{P_j}{M_j}, \\
%\dot P_j=-\frac{\partial H}{\partial R^j}=-\frac{\partial V(R_1,
%  \cdots, R_n)}{\partial R^j}.
%\end{cases}
%\end{equation}
%But these are also the equations corresponding to 
%$$
%\{H,R^j\} =\dot R^j=\frac{P_j}{M_j};\;\{ H, P_j\} =\dot P_j=-\frac{\partial V(R_1,
%  \cdots, R_n)}{\partial R^j}.
%$$
%\end{example}

\section{Summary of geometric quantum mechanics}
\label{sec:append-geom-quant}
The aim of this section is to provide a description of a quantum
mechanical system by using the geometric tools which are used to
describe classical mechanical systems. It is just a very quick
summary of the framework which has been developed in the last 30
years and which can be found in Refs.~\onlinecite{kibble1979geometrization,heslot1985quantum,
abbati1984pure,   cirelli1991quantum,brody2001geometric,Ashtekar:1998p906,
  Carinena:2006p7565,Carinena:2007p813,clemente2008basics} and
references therein.
%Instead of a Hilbert space, we shall be considering the quantum states as
%points of a differentiable manifold $M_Q$. 
For the sake of
simplicity, we shall focus only on the finite dimensional case.
The Hilbert space $\mathcal{H}$ becomes then isomorphic to
$\mathbb{C}^n$ for $n$ a natural number.

\subsection{The states} 
Consider a basis $\{ |\psi_{k}\rangle\}$ for ${\cal H}$. 
Each state $|\psi\rangle\in {\cal H}$ can be written in that basis with complex components  
(or coordinates, in more differential geometric terms)
$\{ z_{k}\}$:
$$
{\cal H}\ni |\psi\rangle=\sum_{k}z_{k}|\psi_{k}\rangle.
$$
We can just take the vector space inherent to the Hilbert
space, and turn it into a real vector space $M_Q$, by splitting each coordinate
in its real and imaginary part:
$$
\mathbb{C}^n\sim \mathcal{H}\ni z_{k}=q_{k}+ip_{k} \mapsto (q_{k}, p_{k})\in 
\mathbb{R}^{2n}\equiv M_{Q}.
$$
We will use real coordinates $(q_k, p_k)$, $k=1, \ldots ,n$ to
represent the points of $\mathcal{H}$ when thought as real manifold
elements. From this real point of view the similarities
between the quantum dynamics and the classical one described 
in the previous section will be more evident.
Sometimes it will be useful to maintain the complex notation $\psi$ 
or $z_k$ for the elements of the Hilbert space. 

Another important aspect of the Hilbert space description of quantum
mechanics is the study of the global phase of the state. It is a well
known fact that physical states are independent from the global phase of
the element of the Hilbert space that we choose to represent
them. In the formulation as a real vector space, we can represent the
multiplication by a phase on the manifold $M_{Q}$  as  a transformation
whose infinitesimal generator is written as:  
\begin{equation}
  \label{eq:phase}
  \Gamma= {i}\sum_{k}\left(  z_{k}\frac {\partial}{\partial z_{k}} -\bar z_{k}\frac {\partial}{\partial \bar z_{k}}\right )
= \sum_{k}\left ( q_{k}\frac {\partial}{\partial p_{k}} -p_{k}
  \frac {\partial}{\partial q_{k}}\right ),
\end{equation}
where the derivatives with respect to the complex and real variables are 
related by 
$\frac {\partial}{\partial z_{k}}=\frac12( \frac {\partial}{\partial q_{k}}
-i \frac {\partial}{\partial p_{k}})
$
and 
$\frac {\partial}{\partial  \bar z_{k}}=\frac12( \frac {\partial}{\partial q_{k}}
+i \frac {\partial}{\partial p_{k}}).
$
The meaning of this vector field is simple to understand if we realize that 
a phase change modifies the angle of the complex number representing the state,
when  considered in polar form 
(i.e., in polar coordinates $\{r_k,
\theta_k\}_{k=1, \cdots , n}$ with $z_k=r_k{\rm e}^{i\theta_k}$, Eq. 
\eqref{eq:phase} becomes
$\Gamma=\sum_k\partial_{\theta_k}$). 
Then, from a geometrical point of view
we can use Eq. \eqref{eq:phase} in two ways: 
\begin{itemize}
  \item Computing its integral curves, which are the different states
    which are obtained from an initial one by a global phase
    multiplication.
  \item Acting with the vector field on functions of $M_Q$ (which will represent our
    observables) providing us with the effect of the global phase
    transformation on the observables. 
\end{itemize}

One final point is to consider the limitation in the norm. The sphere of states with norm equal 
to one in $\mathbb{C}^n$ corresponds in the real-vector-space description to
the $(2N_Q-1)$--dimensional sphere:
\begin{equation}
  \label{eq:32}
  S_Q=\left \{ (\vec q, \vec p)\in M_Q | \sum_k(q_k^2+p_k^2)=1
  \right \}. 
\end{equation}
It is immediate that the vector field \eqref{eq:phase} is tangent
to $S_Q$, since the phase change preserves the norm of the state.
%The vector field 
%\begin{equation}
% \label{eq:33}
%  \Delta=\sum_k \left (q_k\frac{\partial}{\partial q_{k}}+p_{k}\frac
%    {\partial}{\partial p_{k}} \right )=J\Gamma, 
%\end{equation}
%defines the vector field which is normal to the sphere, with respect
%to the Euclidean metric.

% We can  also define the \textbf{complex
%  projective space} as the space where both norm and global phase are
% fixed.  We shall denote this space as $P_Q$.

\subsection{The Hermitian structure}
The Hilbert space structure of a quantum system is encoded in a
Hermitian structure or scalar product $\langle\cdot, \cdot \rangle$.
The latter is specified by the choice of an orthonormal basis
$\{|\phi_k\rangle\}$.

Taking as coordinates $z_k=q_k+ip_k$, 
the components of a vector in this basis, as
we did in the previous paragraph,   
we can define a Poisson bracket in  
$\cal H$, by
\begin{equation}
\label{eq:poisson}
\{f,g\}=2i\sum_k\left(
\frac {\partial f}{\partial z_{k}}\frac {\partial g}{\partial \bar z_{k}}
-
\frac {\partial f}{\partial \bar z_{k}}\frac {\partial g}{\partial z_{k}}
\right)
\end{equation}
or in real coordinates
\begin{equation}
\label{eq:poisson2}
\{f,g\}=\sum_k\left(
\frac {\partial f}{\partial p_{k}}\frac {\partial g}{\partial q_{k}}
-
\frac {\partial f}{\partial q_{k}}\frac {\partial g}{\partial p_{k}}
\right),
\end{equation}
that corresponds to the standar Poisson bracket in classical mechanics.
This justifies the choice of the notation for the real coordinates wich become 
Darboux coordinates for the Poisson bracket.

For later purposes it will be useful to introduce a symmetric bracket
by
\begin{align}
\label{eq:symm}
 \{f,g\}_+=&2\sum_k\left
(\frac {\partial f}{\partial z_{k}}\frac {\partial g}{\partial \bar z_{k}}
+
\frac {\partial f}{\partial \bar z_{k}}\frac {\partial g}{\partial z_{k}}
\right)\\
=&\sum_k
\left(\frac{\partial f}{\partial q_{k}}\frac {\partial g}{\partial q_{k}}
+
\frac {\partial f}{\partial p_{k}}\frac {\partial g}{\partial p_{k}}\right).
\end{align}
Observe that in order to define the Poisson and the symmetric brackets
we made use of an orthonormal basis in the Hilbert space. We may think
that both objects are not canonical as they could depend of the 
choice of the basis. In the following proposition we shall show that this 
is not the case.
% and both brackets are independent of the orthonormal basis.

\begin{proposition}
 The brackets $\{\ ,\ \}$ and   $\{\ ,\ \}_+$ defined above
depend only on the scalar product and not 
on the orthonormal basis used to construct them.
\end{proposition}
\textit{Proof}
The coordinates in two different orthonormal basis, for a 
given scalar product,
$\{|\psi_k\rangle\}$ 
and 
$\{|\psi'_k\rangle\}$ 
are related by a unitary transformation, i.e., if
$$|\psi\rangle=\sum_kz_k|\psi_k\rangle=
\sum_kz'_k|\psi'_k\rangle,$$
we have $z'_j=\sum_kU_j^kz_k$ with
$\sum_kU_j^k\bar U_l^k=\delta_{jl}$.

Therefore
$$
\frac{\partial}{\partial z_k}=
\sum_j U^k_j \frac{\partial}{\partial z'_j}
\quad\text{and}\quad 
\frac{\partial}{\partial \bar z_k}=
\sum_j \bar U^k_j \frac{\partial}{\partial \bar z'_j}.
$$
From which one immediately gets
$$\sum_k \frac{\partial f}{\partial z_k} \frac{\partial g}{\partial \bar z_k}=
\sum_j \frac{\partial f}{\partial z'_j} \frac{\partial g}{\partial \bar z'_j}
$$
and both the Poisson and symmetric brackets are independent
from the choice of orthonormal basis, which concludes the proof.
\qed

An important property of the brackets defined above is that
they are preserved by the vector field $\Gamma$ in
Eq. (\ref{eq:phase}) in the sense that
$$
\Gamma\{f,g\}= \{\Gamma f,g\}+\{f,\Gamma g\}
$$
and
$$
\Gamma\{f,g\}_+= \{\Gamma f,g\}_++\{f,\Gamma g\}_+.
$$
A fact that can be proved with a simple computation.

This property is important for us because it implies
that the symmetric or antisymmetric bracket of two functions
killed by $\Gamma$ (i.e., constant under phase change)
will be killed by $\Gamma$ too:
$$
\Gamma f=\Gamma g=0\Rightarrow 
\Gamma\{f,g\}
=
\Gamma\{f,g\}_+=0.
$$
Therefore, it makes sense to restrict ourselves to this class 
of functions.

\subsection{The observables}
\label{sec:observ}

Now that we have been able to formulate quantum mechanics on a real
state space, we proceed to discuss how to represent the physical
observables on this new setting. Instead of considering the
observables as linear operators (plus the usual requirements,
self-adjointness, boundedness, etc.) on the Hilbert space $\mathcal{H}$, we
shall be representing them as functions defined on the 
real manifold $M_Q$. The reason for that is to resemble, as much
as possible, the classical mechanical approach. But we can not forget
the linearity of the operators, and thus the functions must be chosen
in a very particular way. The usual choice inspired in Ehrenfest's theorem
selects the functions introduced in the following:
\begin{definition}
\label{sec:observables}
  To any operator $A\in \mathrm{Lin}(\mathcal{H})$ we
  associate the quadratic function
  \begin{equation}
    \label{eq:2}
    f_A(\psi)=\frac 12\langle\psi|A\psi\rangle.
\end{equation}
We shall denote  the set of such functions as $\mathcal{F}_s(M_Q)$.
\end{definition}
%\begin{note}
%  The term bilinear is used here in some special
%  sense. Bilinearity is usually defined as a property of bilinear
%  forms on   real vector spaces, when considering scalar products, for
%  instance. The representation above uses indeed a bilinear
%  bilinear form, which corresponds to the Hermitian operator $A$
%  written in terms of the real-vector space structure of $M_Q$.  But
%  we use just its 'diagonal' part, since we just evaluate the
 % form on $(\psi,\psi)$. But we prefer to keep the term bilinear
  %for we consider it provides the best description of the nature of
  %the function.
%\end{note}

Notice that this definition of observable is different from
the analogous one in the classical case. In the classical framework a
state, represented by a point in $M_Q$, provides a well defined result
for any observable $f:M_C\to \mathbb{R}$. In this  geometric quantum mechanics 
(GQM in the following) framework, on the
other hand, the value at a given state of  $f_A\in
\mathcal{F}_s(M_Q)$ provides just the average value of the corresponding
operator in that state. Besides, contrarily to the classical case,
not all the $C^\infty$ functions on $M_Q$ are regarded as observables, but
only those of the form of \eqref{eq:2}.
%This is implementing, in a geometric
%language, the probabilistic formulation which is encoded in Quantum
%Mechanics.

Once the definition has been stated, we must verify that it is a
consistent one, in the sense that the brackets defined in $\cal H$
act correctly in this space. This is contained in the following 
 
\begin{lemma}
  The set of functions of the form \eqref{eq:2} is closed under the
  brackets defined in section B.
\end{lemma}
\textit{Proof}
  Direct computation: just take the expression of the functions, compute the derivatives, obtain the bracket and verify that they correspond 
to the  expressions of the functions associated to the 
(anti)commutator of the corresponding operators.\qed

Within the usual approach of quantum mechanics, there
are three algebraic structures on the set of operators, which turn
out to be meaningful and important for the physical description:
\begin{itemize}
\item The associative product of two operators: 
$$
A, B\in  \mathrm{Lin}(\mathcal{H}) \to A.B \in \mathrm{Lin}(\mathcal{H}).
$$
It is important to notice, though, that this operation is not internal in
the set of Hermitian operators (i.e., those associated to physical magnitudes), since the 
product of two Hermitian operators is not Hermitian, in general. 
\item The anticommutator of two operators:
$$
A, B\in  \mathrm{Lin}(\mathcal{H}) \to [A,B]_+ \in \mathrm{Lin}(\mathcal{H}).
$$
\item The commutator of two operators:
$$
A, B\in  \mathrm{Lin}(\mathcal{H}) \to i[A,B] \in \mathrm{Lin}(\mathcal{H}).
$$
\end{itemize}
Notice that these last two operations are internal in the space of Hermitian operators.
How do we translate these operations into the GQM scheme? 

To answer this question we recall the brackets defined in the previous 
subsection B.
If we take  $f_A, f_B\in  \mathcal{F}_s(M_Q)$,
\begin{itemize}
%We shall come back to this structure at the end of the paper.
\item the anticommutator of two operators, becomes the symmetric bracket
or Jordan product of the functions 
(see Ref. \onlinecite{landsnab}):
$$
[A, B]_{+}\to f_{[A,B]_+}=\{f_A, f_B\}_+ \in \mathcal{F}_s(M_Q).
$$
\item The commutator of two operators, transkates into the Poisson bracket of
  the functions:
$$
i[A,B]\to f_{i[A,B]}=\{f_A, f_B\} \in \mathcal{F}_s(M_Q).
$$
\end{itemize}

\textbf{Conclusion}: The operators and their algebraic structures are
encoded in the set of functions with the brackets associated
to the Hermitian product.

Thus we see why it makes sense to consider only that specific type of
functions: it is a choice which guarantees to maintain all the
algebraic structures which are required in the quantum description.

Another important property of the set of operators of quantum mechanics is the 
corresponding spectral theory.  In any quantum system, it is of the utmost importance to be 
able to find  eigenvalues and eigenvectors. We can summarize the
relation between these objects and those in the new GQM scheme in the
following result: 
\begin{lemma}
  Let $f_{A}$ be the function associated to the observable $A$. Then, if we consider 
$\psi\neq 0$ (The case $\psi=0$ is not meaningful from the physical point of view),
  \begin{itemize}
  \item the eigenvectors of the operator $A$ coincide with the critical points of the function $f_{A}$, i.e.,
$$
df_{A}(\psi)=0 \Rightarrow \psi \text{ is an eigenvector of } A.
$$
\item The eigenvalue of $A$ at the eigenvector $\psi$ is the value that the function $f_{A}$ takes at the critical point $\psi$.
  \end{itemize}
\end{lemma}
\textit{Proof}
  Trivial computation. Recall Ritz theorem from any quantum mechanics textbook, as
for instance Ref.~\onlinecite{cohen:1973}.\qed

In the previous subsection we saw that the vector field defined in
Eq.(\ref{eq:phase})\cite{Carinena:2006p7565} preserves the Poisson 
bracket. Indeed, it is possible to
prove that $\Gamma$ corresponds to a Hamiltonian vector field whose
Hamiltonian function is precisely the norm of the state: 
\begin{lemma}
\label{lemma_norma}
  Let $M_{Q}=\mathbb{R}^{2n}$. The vector field defining the
  multiplication of the states by a phase is the Hamiltonian vector
  field of the function corresponding to the identity operator: 
  \begin{equation}
    \label{eq:phasevec}
f_\mathbb{I}=\frac 12\langle\psi | \mathbb{I}\psi \rangle\longrightarrow X_{f_\mathbb{I}}=\Gamma.    
  \end{equation}
\end{lemma}
\textit{Proof}
  The function $f_\mathbb{I}$ is easily computed:
$$
f_\mathbb{I}=\frac 12 \sum_{k}\left(q_{k}^{2}+p_{k}^{2}\right ).
$$
The Hamiltonian vector field corresponding to this function is, easily too, 
computed: 
$$
\{ f_\mathbb{I}, \cdot \}=\sum_k \left (p_{k}\frac{\partial} 
{\partial q_{k}}- q_{k}\frac{\partial}{\partial p_{k}}\right )=\Gamma.
$$\qed

\subsection{The dynamics}

As in the classical case, the dynamics can be implemented in different forms, always in a way
which is compatible with the geometric structures introduced so far:
\begin{itemize}
\item  In the Schr{\"{o}}dinger picture, the dynamics is described as the integral curves
  of the vector field $X_{f_H}$, where $H$ is the Hamiltonian operator of the system:
  \begin{equation}
    \label{eq:4}
    X_{f_H}=\hbar^{-1}\{  f_H, \cdot \}.
  \end{equation}

  %In the Schr\"odinger picture, dynamics is described as the
  %integral curves of a vector field $X_H$ (with $H$ the Hamiltonian
  %operator of the system), which turns out to be the
  %Hamiltonian vector field of the function $f_H$ times $\hbar^{-1}$:
  %\begin{equation}
  %  \label{eq:4}
  %  X_H=\hbar^{-1}\{  f_H, \cdot \}.
  %\end{equation}

\item In the Heisenberg picture, the dynamics is introduced by
  transferring the Heisenberg equation into the language of functions:
  \begin{equation}
    \label{eq:3}
    \dot f_A=\hbar^{-1}\{f_H, f_A \}.
  \end{equation}
\end{itemize}

The most interesting example is probably the expression of the von Neumann
equation:
\begin{equation}
  \label{eq:5}
   \dot f_\rho=\hbar^{-1}\{ f_H, f_\rho \},
\end{equation}
where $f_\rho$ is the function associated to the density matrix.

\textbf{Conclusion}: It is possible to describe quantum dynamics as a
flow of a vector field on the manifold $M_Q$ or on the set of quadratic
functions of the manifold. Such a vector field is a Hamiltonian vector
field with respect to the Poisson bracket encoded in the
Hermitian structure of the system.

\begin{example}
\label{sec:dynamics}
  Let us consider again the simplest quantum situation defined on
  $\mathbb{C}^2$. As a real manifold, $M_Q\sim \mathbb{R}^4$. Consider
  then a Hamiltonian $H:\mathbb{C}^2\to\mathbb{C}^2$ which is usually
  written as a matrix:
$$
H=
\begin{pmatrix}
  H_{11} & H_{12} \\
H_{21} & H_{22}
\end{pmatrix}.
$$  
If we consider it as a matrix on the real vector space $M_Q$, it
reads:
$$
H_{\mathbb{R}}=
\begin{pmatrix}
  H_{q_1q_1} & H_{q_1p_1} &H_{q_1q_2} & H_{q_1p_2}\\
  H_{p_1q_1} & H_{p_1p_1} &H_{p_1q_2} & H_{p_1p_2}\\
  H_{q_2q_1} & H_{q_2p_1} &H_{q_2q_2} & H_{q_2p_2}\\
  H_{p_2q_1} & H_{p_2p_1} &H_{p_2q_2} & H_{p_2p_2}
\end{pmatrix},
$$  
where the matrix above is symmetric because $H$ is Hermitian, since we
have: 
\begin{equation*}
\begin{gathered}[c]
H_{q_1q_1}=H_{11}=H_{p_1p_1},
\\
H_{q_2q_2}=H_{22}=H_{p_2p_2},
\\
H_{q_1q_2}=\mathrm{Re}(H_{12})=H_{p_1p_2},
\\
H_{q_2q_1}=\mathrm{Re}(H_{21})=H_{p_2p_1},
\end{gathered}
\;
\begin{gathered}[c]
H_{q_1p_1}=0=H_{p_1q_1},
\\
H_{q_2p_2}=0=H_{p_2q_2},
\\
H_{q_1p_2}=-\mathrm{Im}(H_{12})=-H_{p_1q_2},
\\
H_{q_2p_1}=-\mathrm{Im}(H_{21})=-H_{p_2q_1}.
\end{gathered}
\end{equation*}

The function $f_H$ in $\mathcal{F}_s(M_Q)$ becomes thus:
$$
f_H=\frac{1}{2}\psi_{\mathbb{R}}^{t}H_{\mathbb{R}}\psi_{\mathbb{R}},
\quad\text{where}\quad \psi_{\mathbb{R}}=(q_{1},p_1,q_{2},p_2)^{t}
$$
% $$
% f_H= 
% \frac 12\begin{pmatrix}
%   q_{1},  p_1,  q_{2},  p_2 
% \end{pmatrix}
% \begin{pmatrix}
%   H_{q_1q_1} & H_{q_1p_1} &H_{q_1q_2} & H_{q_1p_2}\\
%   H_{p_1q_1} & H_{p_1p_1} &H_{p_1q_2} & H_{p_1p_2}\\
%   H_{q_2q_1} & H_{q_2p_1} &H_{q_2q_2} & H_{q_2p_2}\\
%   H_{p_2q_1} & H_{p_2p_1} &H_{p_2q_2} & H_{p_2p_2}
% \end{pmatrix}
% \begin{pmatrix}
%  q_{1} \\
% p_1 \\
% q_{2} \\
% p_2 
% \end{pmatrix},
% $$
% where the matrix above is symmetric because $H$ is Hermitian, since we
% have: 
% \begin{equation*}
% \begin{gathered}[c]
% H_{q_1q_1}=H_{11}=H_{p_1p_1},
% \\
% H_{q_2q_2}=H_{22}=H_{p_2p_2},
% \\
% H_{q_1q_2}=\mathrm{Re}(H_{12})=H_{p_1p_2},
% \\
% H_{q_2q_1}=\mathrm{Re}(H_{21})=H_{p_2p_1},
% \end{gathered}
% \;
% \begin{gathered}[c]
% H_{q_1p_1}=0=H_{p_1q_1},
% \\
% H_{q_2p_2}=0=H_{p_2q_2},
% \\
% H_{q_1p_2}=-\mathrm{Im}(H_{12})=-H_{p_1q_2},
% \\
% H_{q_2p_1}=-\mathrm{Im}(H_{21})=-H_{p_2q_1}.
% \end{gathered}
% \end{equation*}
% 
% \begin{equation*}
% \begin{gathered}[c]
% H_{q_1q_1}=H_{11}=H_{p_1p_1},
% \\
% H_{q_2q_2}=H_{22}=H_{p_2p_2},
% \end{gathered}
% \qquad
% \begin{gathered}[c]
% H_{q_1p_1}=0=H_{p_1q_1},
% \\
% H_{q_2p_2}=0=H_{p_2q_2},
% \end{gathered}
% \end{equation*}
% \begin{equation*}
% \begin{gathered}[c]
% H_{q_1q_2}=\mathrm{Re}(H_{12})=H_{p_1p_2},
% \\
% H_{q_2q_1}=\mathrm{Re}(H_{21})=H_{p_2p_1};
% \end{gathered}
% \quad
% \begin{gathered}[c]
% H_{q_1p_2}=-\mathrm{Im}(H_{12})=-H_{p_1q_2},
% \\
% H_{q_2p_1}=-\mathrm{Im}(H_{21})=-H_{p_2q_1}.
% \end{gathered}
% \end{equation*}
and then, the Hamiltonian vector field turns out to be:
$$
  X_H=\hbar^{-1}\left ( \frac{\partial f_H}{\partial p_1}\frac{\partial}{\partial q_{1}}
- \frac{\partial f_H}{\partial q_{1}}\frac{\partial}{\partial p_1}+
\frac{\partial f_H}{\partial p_2}\frac{\partial}{\partial q_{2}}
-\frac{\partial f_H}{\partial q_{2}}\frac{\partial}{\partial p_2} \right )
$$
and its integral curves are precisely the expression of Schr\"odinger
equation when we write it back in complex terms:
\begin{eqnarray*}
    \dot q_{1}&=&\hbar^{-1}(H_{p_1q_1}q_{1}+H_{p_1q_2}q_{2}+H_{p_1p_1}p_1+H_{p_1p_2}p_2), 
    \\ 
    \dot p_1&=-&\hbar^{-1}(H_{q_1q_1}q_{1}+H_{q_1q_2}q_{2}+H_{q_1p_1}p_1+H_{q_1p_2}p_2),    
    \\
    \dot q_{2}&=&\hbar^{-1}(H_{p_2q_2}q_{2}+H_{p_2q_1}q_{1}+H_{p_2p_2}p_2+H_{p_2p_1}p_1),
    \\ 
    \dot p_2&=-&\hbar^{-1}(H_{q_2q_2}q_{2}+H_{q_2q_1}q_{1}+H_{q_2p_2}p_2+H_{q_2p_1}p_1).
\end{eqnarray*}

% \begin{equation*}
%   \begin{cases}
%     \dot q_{1}= \hbar^{-1}(H_{p_1q_1}q_{1}+H_{p_1q_2}q_{2}+H_{p_1p_1}p_1+H_{p_1p_2}p_2) \\ 
%     \dot p_1=-\hbar^{-1}(H_{q_1q_1}q_{1}+H_{q_1q_2}q_{2}+H_{q_1p_1}p_1+H_{q_1p_2}p_2
% )    \\
%      \dot q_{2}=\hbar^{-1}(H_{p_2q_2}q_{2}+H_{p_2q_1}q_{1}+H_{p_2p_2}p_2+H_{p_2p_1}p_1)\\ 
%     \dot p_2=-\hbar^{-1}(H_{q_2q_2}q_{2}+H_{q_2q_1}q_{1}+H_{q_2p_2}p_2+H_{q_2p_1}p_1)
%   \end{cases}
% \end{equation*}

We can write these equations as:
\begin{equation}
  \label{eq:realsch}
\dot{\psi_{\mathbb{R}}}=-\hbar^{-1} \mathbf{J}
H_{\mathbb{R}}\psi_{\mathbb{R}}
\end{equation}
% 
% \begin{equation}
%   \label{eq:realsch}
%   \begin{pmatrix}
%     \dot q_{1}\\
% \dot p_1 \\
% \dot q_{2} \\
% \dot p_2 
%   \end{pmatrix}=-\hbar^{-1}
%  \mathbf{J}
% \begin{pmatrix}
%   H_{q_1q_1} & H_{q_1p_1} &H_{q_1q_2} & H_{q_1p_2}\\
%   H_{p_1q_1} & H_{p_1p_1} &H_{p_1q_2} & H_{p_1p_2}\\
%   H_{q_2q_1} & H_{q_2p_1} &H_{q_2q_2} & H_{q_2p_2}\\
%   H_{p_2q_1} & H_{p_2p_1} &H_{p_2q_2} & H_{p_2p_2}
% \end{pmatrix}
% \begin{pmatrix}
%  q_{1}\\
% p_1 \\
% q_{2} \\
% p_2 
% \end{pmatrix},
% \end{equation}
with
$$\mathbf{J}= 
   \begin{pmatrix}
     0 & -1 & 0 &  0 \\
     1 & 0 & 0 &0 \\
     0& 0 & 0 & -1 \\
     0 & 0 & 1 & 0
    \end{pmatrix};
$$
or, equivalently, 

\begin{equation}
  \label{eq:7}
 |\dot\psi\rangle=-\hbar^{-1} i H|\psi\rangle  
\end{equation}
 where
\begin{equation}
  \label{eq:26}
  |\psi\rangle=
\begin{pmatrix}
 q_{1} + i p_1 \\
q_{2} + i p_2 
\end{pmatrix}
\end{equation}
is the complex state vector in terms of the real coordinates. 
The operator $\mathbf{J}$, that satisfies $\mathbf{J}^2=-I$,
is called the complex structure. As it is apparent when comparing
(\ref{eq:realsch}) and (\ref{eq:7}), it
implements the multiplication by the imaginary unity $i$ 
in the real presentation of the Hilbert space. 
Observe that $\mathbf{J}$ and $H_{\mathbb{R}}$ commute, this is due to the
fact that the latter comes from an operator $H$ in the complex Hilbert space.
\end{example}

As the norm of a state can be written  as the function associated to
the identity operator, it is trivial to prove: 
\begin{lemma}
\label{lemma:norma}
  The dynamics defined by Eqs (\ref{eq:3}) (equivalently by
  (\ref{eq:4})) preserves  the norm of the state.
\end{lemma}
\textit{Proof}
  It is immediate to see that the evolution of the function
  $f_{\mathbb{I}}$ is given by the Poisson bracket with the
  Hamiltonian: 
$$
\dot f_{\mathbb{I}}=\hbar ^{-1}\{ f_{H}, f_{\mathbb{I}} \}.
$$
But because of the properties of the bracket 
$$
\{f_H , f_{\mathbb{I}} \}=f_{[H,\mathbb{I}]}=0.
$$
Thus the dynamics preserves the norm of the state, and the flow is
restricted to the sphere $S_Q$.\qed
%%\end{proof}

\section{Ehrenfest dynamics as a Hamiltonian system}
\label{sec:our-proposal}
%Let us summarize the most relevant consequences of the previous
%section.  
%It is important to remark that the construction is intrinsic and
%  tensorial, and thus it is automatically coordinate-independent.
In this section we show how to put together 
the dynamics of a quantum and a classical system 
following the presentation of the two previous sections.
We are describing thus a physical system characterized by the
following elements:

\subsection{The set of states of our system}
\begin{itemize}
 \item First, let $\mathcal{H}$ be  a Hilbert space  which
   describes the quantum degrees of freedom of our system.
   For example, it could describe the electronic subsystem; in this case, 
   it is the vector space corresponding to the completely
   antisymmetric representation of the permutation group $S_{N}$
   (i.e., a set of Slater determinants), where $N$ is the number of
   electrons of the system and each electron lives in a Hilbert space
   of dimension $M$. Thus, the dimension of ${\cal H}$ will be
   $N_{Q}=\binom{M}{N}$.

   We know that it is a complex vector space, but we prefer to consider it
   as a real vector space with the double of degrees of freedom and
   denote it as $M_Q$. Also, in correspondence with the Hilbert space
   vectors in the usual formalism of quantum mechanics, several states
   in $M_Q$ represent the same physical state. To consider true physical states one should 
   extract only those corresponding to the projective space, which can
   be identified with a submanifold of $M_Q$. A more general approach
   is to consider the sphere of states with norm equal to one, $S_Q$, and
   take into account  the phase transformations generated by Eq
   \eqref{eq:phase} in a proper way. We will discuss this in the following sections.

% if we consider the canonical Poisson bracket associated with
%  the canonical symplectic tensor of the function 
%$$
%f_H=\frac 12 \langle\psi, H\psi\rangle
%$$
 \item Second, let $M_C$ be  a differentiable manifold which
   contains the classical degrees of freedom. We shall assume it to
   be a phase space, and thus it will have an even number of degrees of
   freedom and it shall be endowed with a non degenerate Poisson bracket
   that in Darboux cordinates $(\vec R,\vec P)$ reads
   $$\{f,g\}_C=\sum_J \frac{\partial f}{\partial P_J}\frac{\partial g}{\partial R_J}-\frac{\partial f}{\partial R_J}\frac{\partial g}{\partial P_J}.$$

 \item  Third, we let our state space $\mathcal{S}$ be the Cartesian
   product of both manifolds, 
   $$
   \mathcal{S}=M_C\times M_{Q}.
   $$
   Such a description has important implications: it is possible to consider each subsystem
   separately in a proper way but it is not possible to entangle the
   subsystems with one another. As long as Ehrenfest dynamics
   disregards this possibility, the choice of the Cartesian product is
   the most natural one.
\end{itemize}

\begin{example}
If we consider a simple case, where we have one nucleus moving in a
three dimensional domain and the electron state is considered to belong 
to a two-level system, the situation would be:
\begin{equation*}
  \Psi=(\vec R, \vec P,q_1, q_2, p_1, p_2),\qquad (q_1, q_2, p_1, p_2)\in \mathbb{R}^4,
\end{equation*}
where $\vec R$ represents the position of the nucleus, and $\vec P$
represents its linear momentum. The tetrad $(q_1, q_2, p_1, p_2)$
represents the set of four real coordinates which correspond to the
representation of the state of the two-level system on a real vector
space (of four 'real' dimensions which corresponds to a
'complex'  two-dimensional vector space).  
\end{example}

As a conclusion from the example above, we use as
coordinates for our states:
\begin{itemize}
\item The positions and momenta of the nuclei: 
  \begin{equation}
    \label{eq:11}
    (\vec R, \vec P)\in M_C.
  \end{equation}
We will have $3N_C+3N_C$ of these, for $N_C$ the number of nuclei of the
system.  
\item The real and imaginary parts of the coordinates of the Hilbert
  space elements with respect to some basis:
  \begin{equation}
    \label{eq:12}
    (\vec q, \vec p)\in M_Q.
  \end{equation}
We will have $N_Q+N_Q$ of these, for $N_Q$ the complex dimension of the
Hilbert space ${\cal H}$.
\end{itemize}

\subsection{The observables}
To represent the physical magnitudes we must consider also 
the classical-quantum observables from a new perspective. 

Our observables must be functions defined on the state space
$S=M_C\times M_Q$. 
%We can consider also the projections:
%\begin{align}
%  \label{eq:1}
%  &\pi_C:M_C\times M_Q\to M_C, \quad \pi_C(\vec R, \vec P,\vec q, \vec
%  p)=(\vec R, \vec P)\\
% \intertext{and}
% \label{eq:14}
%  &\pi_Q:M_C\times M_Q\to M_Q, \quad \pi_Q(\vec R, \vec P,\vec q, \vec
%  p)=(\vec q, \vec p).
%\end{align}
We know from our discussion in the case of a purely quantum system
that any function of the  form  (\ref{eq:2}) produces an evolution,
via the Poisson bracket, which preserves the norm. In the MQCD case, we
can easily write the analogue of the vector field (\ref{eq:phase}) by
writting: 
\begin{equation}
  \label{eq:17}
  \Gamma_{Q}=\mathbb{I}\otimes \Gamma.
\end{equation}
%It is simple to see that this object is completely determined by the pullback of the 
%projections $\pi_{C}$ and $\pi_{Q}$ (see below for an explicit definition):
%$$
%\Gamma_{Q}\pi_{C}^{*}=0, \qquad \Gamma_{Q}\pi_{Q}^{*}=\Gamma.
%$$

This is again the infinitesimal generator of phase transformations for
the quantum subsystem, but written at the level of the global state
space $M_{C}\times M_{Q}$.  A reasonable property to be asked to the
functions chosen to represent our observables is to be
constant under this transformation. From a mathematical point of view we can
write such a condition as follows: 

% **************************************

\begin{definition}
We shall define the set of possible observables, $\mathcal{O}$, as the set of all 
$C^\infty$--functions on the set $M_C\times M_Q$ which are constant
under phase changes on the quantum degrees of freedom  ,i.e., 
\begin{equation}
  \label{eq:10}
  \mathcal{O}=\{ f\in C^\infty (M_C\times M_Q) | \, \, \Gamma_{Q}f=0 \}.
\end{equation}

\end{definition}

As we shall see later, this choice reflects the fact that, when considered coupled
together, the nonlinearity of classical mechanics expands also to MQCD. 

We would like to remark that because of the choice of the set of
states as a Cartesian product of the classical states and the quantum
states, we can consider as subsets of the set of observables:
\begin{itemize}
\item The set of classical functions: these are functions which depend
  only on the classical degrees of freedom. Mathematically, they can be
  written as those functions $f\in \mathcal{O}$ such that there exists
  a function $f_C\in C^\infty(M_C)$ such that
%$$
%f=\pi_C^*(f_C); \text{ i.e. } f(\vec R, \vec P, \vec q, \vec
%p)=f_C(\vec R, \vec P),
%$$
$$
f(\vec R, \vec P, \vec q, \vec
p)=f_C(\vec R, \vec P)\,.
$$
%for $\pi_C^*$ the pullback of the projection $\pi_C$. 
We  denote
this subset as $\mathcal{O}_C$. An example of a function belonging
to this set is the linear momentum of the nuclei.

\item the set of \textbf{generalized quantum functions}: functions which depend
  only on the quantum degrees of freedom and which are constant under
  changes in the global phase.  Mathematically, they can be 
  written as those functions $f\in \mathcal{O}$ such that there exists
  a function $f_Q\in C^\infty(M_Q)$ such that
%  \begin{equation}
%    \label{eq:15a}
%    f=\pi_Q^*(f_Q);\text{ i.e. } f(\vec R, \vec P, \vec q, \vec
%p)=f_Q(\vec q, \vec p); \quad \Gamma(f_{Q})=0
%  \end{equation}
  \begin{equation}
    \label{eq:15a}
    f(\vec R, \vec P, \vec q, \vec
p)=f_Q(\vec q, \vec p); \quad \Gamma(f_{Q})=0\,.
  \end{equation}
%for $\pi_Q^*$ the pullback of the projection $\pi_Q$. 
We  denote these functions as $\mathcal{O}_Q$.  We have added
the adjective ``generalized'' because this set is too large to
represent only the set of pure quantum observables. These later functions, should
be considered, when necessary, as a smaller subset, which corresponds to
the set of functions defined in Eq. \eqref{eq:2}. We
denote this smaller subset as $\mathcal{O}_{Q}^s$. An example of
a function belonging to $\mathcal{O}_{Q}^s$ is the linear momentum of the electrons. 
\item A third interesting subset is the set of arbitrary linear
  combinations of the subsets above, i.e., those functions which are
  written as  the sum of a purely classical function and a purely
  quantum one:
%\begin{align}
%    \label{eq:15b}
%    f=&\pi_Q^*(f_Q)+\pi_C^*(f_C);\text{ i.e. } \nonumber \\ 
%f(\vec R, \vec P, \vec q, \vec
%p)=&f_C(\vec R, \vec P)+f_Q(\vec q, \vec p)
%  \end{align}
\begin{align}
  \label{eq:15b}
  f(\vec R, \vec P, \vec q, \vec
  p)=&f_C(\vec R, \vec P)+f_Q(\vec q, \vec p).
\end{align}
We will denote this set as $\mathcal{O}_{C+Q}$. An element of this set
of functions is the total linear momentum of the composed system.
\end{itemize}

We would like to make a final but very important remark. We have not
chosen the set of observables as
\begin{equation}{\cal P}=
\label{eq:42}
 \left \{ f_{A}\in C^\infty(M_C\times M_Q) |
    f_{A}=\langle\psi|
    A(\vec R, \vec P) \psi\rangle \right \}, 
\end{equation}
for $A(\vec R, \vec P)$ a linear operator on the Hilbert space
$\mathcal{H}$ depending on the classical degrees of freedom because of
two reasons:
\begin{itemize}
\item It is evident that the set above is a subset of \eqref{eq:10}
  and thus we are not loosing any of these observables. 
But it is a well known property that Ehrenfest dynamics is
  not linear and then if we consider the operator describing the
  evolution of the system, it can not belong to the set above. We must
  thus enlarge the set \eqref{eq:42}. 
\item  We are going to introduce in
  the next section a Poisson bracket on the space of observables. For
  that bracket to close a Poisson algebra, we need to consider the
  whole set \eqref{eq:10}. 
\end{itemize}

It is important to notice that in the set \eqref{eq:10} there are
operators which are not representing linear operators for the quantum
part of the system and hence the set of properties listed in Section
\ref{sec:append-geom-quant} for the pure quantum case are meaningless
for them. But this is a natural feature of the dynamics we are
considering, because of its nonlinear nature.

\subsection{Geometry and the Poisson bracket on the classical-quantum world} 
\label{sec:dynam-class-quant}

Finally, we must combine the quantum and the classical description in
order to provide a unified description of our system of interest. As
we assume that both the classical and the quantum subsystems are
endowed with Poisson brackets, we face the same problem we have when
combining, from a classical mechanics perspective, two classical
systems.  Therefore it is immediate to conclude that the corresponding
Poisson structures can  be combined as: 
\begin{equation}
  \label{eq:20}
\{ \cdot, \cdot\}= \{ \cdot, \cdot\}_C+\hbar^{-1}\{ \cdot, \cdot\}_Q,
\end{equation}
where the term $ \{ \cdot, \cdot\}_C$ acts on the degrees of freedom
of the first manifold and  $\{ \cdot, \cdot\}_Q$ acts on the degrees
of freedom of the second one. It is a known fact\cite{Abraham:1978p1131} that 
such a superposition of Poisson brackets 
always produces a well defined Poisson 
structure in the product space.

% \begin{remark}
The set of pure classical functions $\mathcal{O}_C$ and the set of
quantum generalized functions $\mathcal{O}_Q$  are closed under the Poisson
bracket.  The same happens with the quantum functions
$\mathcal{O}_Q^s$ and the set of linear combinations
$\mathcal{O}_{C+Q}$. In mathematical terms, what we have is a family of
Poisson subalgebras. This property ensures that the description of
purely classical or purely quantum systems, or even both systems at
once but uncoupled to each other, can be done within the formalism. 
% \end{remark}

Once the Poisson bracket on $M_{C}\times M_{Q}$ has been  introduced
we can express again the constraint we introduced in the definition of
the observables in Poisson terms.  Thus we find that in a completely
analogous way to the pure quantum case, we can prove that 
\begin{lemma}
\label{sec:geom-poiss-brack}
 The condition in Eq. \eqref{eq:10}
$$
\Gamma_{Q}(f)=0 
$$
is equivalent to ask the function $f$ to Poisson-commute with the
function $f_{\mathbb{I}}=\sum_{k}(q_{k}^{2}+p_{k}^{2})$, i.e., 
$$
\Gamma_{Q}(f)=0 \Leftrightarrow \{ f_{\mathbb{I}}, f\}=0.
$$
\end{lemma}

\subsection{The definition of the dynamics}

From  Sections \ref{sec:append-geom-dynam-PB} and
\ref{sec:append-geom-quant}, our
formulation of MQCD can be implemented on:
\begin{itemize}
\item The manifold which represents the set of states by defining a
  vector field whose integral curves represent the solutions of the
  dynamics (Schr\"odinger picture).
\item The set of functions (please note the differences between the
  classical and the quantum cases) defined on the set of states which
  represent the set of observables of the system. In this case the
  Poisson bracket of the functions with the Hamiltonian of the system
  defines the corresponding evolution (Heisenberg picture).
\end{itemize}

Both approaches are not disconnected, since they can be easily
related:
\begin{equation}
  \label{eq:18}
  X_H=\{ f_H,\cdot \}, 
\end{equation}
where we denote by $X_H$ the vector field which represents the
dynamics on the phase space and by $f_H$ the function which
corresponds to the Hamiltonian of the complete system.

We can now proceed to our first goal: to provide a Hamiltonian
description of Ehrenfest dynamics in terms of a Poisson structure. 
We thus define the following Hamiltonian system: 
\begin{itemize}
  \item A state space corresponding to the Cartesian product
        $M_C\times M_Q$.

  \item A set of operators corresponding to the set of functions
        $\mathcal{O}$ defined in Eq. \eqref{eq:10}.
        
  \item The Poisson bracket  defined in Eq. \eqref{eq:20}.
        
  \item And finally, the dynamics introduced by the following
        Hamiltonian function: 
        \begin{equation}
         \label{eq:9}
         f_H(\vec R, \vec P,\vec q, \vec p)=
         \sum_{J}\frac{\vec P_{J}^2}{2M_{J}}+\langle
         \psi(\vec q, \vec p)| H_e( \vec R) \psi(\vec q, \vec p)\rangle,
        \end{equation}
        where $H_e$ is the expression of the electronic Hamiltonian,  $M_J$
        are the masses of the classical subsystem of the nuclei and $\psi( \vec
        q, \vec p)$ is the real-space representation of the state $\psi$
        analogous to Eq. \eqref{eq:26}.        
\end{itemize}

 As a result, the dynamics of both subsystems are
obtained easily. In the
Schr\"odinger picture we obtain: 
\begin{align}
  \label{eq:6}
\dot{\vec R}&= \frac{\partial f_H}{\partial \vec P}=M^{-1}\vec P, 
\\
  \label{eq:6-2}
\dot{\vec P}&= -\frac{\partial f_H}{\partial \vec R}=-
\mathrm{grad}(\langle \psi(\vec q,\vec p)| H_e(\vec R) \psi(\vec q,
\vec p)\rangle), 
\\ 
%\dot  \psi= -i H_e(\vec R)\psi
  \label{eq:6-3}
\dot q_1&= \hbar^{-1}\frac{\partial f_H}{\partial p_1}, 
\\
  \label{eq:6-4}
\dot p_1&= - \hbar^{-1}\frac{\partial f_H}{\partial q_1}, 
\\
\notag
&\vdots 
\\
  \label{eq:6-5}
\dot q_{N_Q}&= \hbar^{-1}\frac{\partial f_H}{\partial p_{N_Q}}, 
\\
  \label{eq:6-6}
\dot p_{N_Q}&= -\hbar^{-1}\frac{\partial f_H}{\partial q_{N_Q}}.
\end{align}
%where the last equations correspond to the Schr\"odinger equation
%written for the simple case of $M_Q=\mathbb{R}^{2n}$. 
Remark that, form a formal point of view, this is analogous to Example
\ref{sec:dynamics}.  This set of
equations corresponds exactly with Ehrenfest dynamics.

The final point is to prove the following lemma:
\begin{lemma}
  The dynamics preserves the set of observables $\mathcal{O}$.
\end{lemma}
\textit{Proof}
 From Lemma \ref{lemma:norma}, an observable belongs to $\mathcal{O}$
 if it Poisson-commutes with 
  $f_{\mathbb{I}}$. Thus, as $f_H\in \mathcal{O}$, if we consider an
  observable $f\in \mathcal{O}$,  by the Jacobi identity:
  \begin{equation}
\{ f_{\mathbb{I}},\{f_H, f\}\} =-\{f,\{f_{\mathbb{I}},
    f_H\}\}-\{f_H,\{f, f_{\mathbb{I}}\}\}=0.
\end{equation}
\qed
%\end{proof}\qed

%This is equivalent to the claim:
%\begin{lemma}
%  If we take an initial condition on $M_C\times S_Q$ (i.e. a quantum
%  state with norm equal to one) or in $M_C\times P_Q$ the dynamics
%  stays on that manifold. 
%\end{lemma}

\section{Statistics of the Ehrenfest dynamics}
\label{sec:defin-stat-syst}

The next step in our work is  the definition of a statistical system
associated to the dynamics we introduced above. The first ingredient
for that is the definition of the distributions we shall be describing
the system with.

The main conclusion from the previous section is that Ehrenfest dynamics
can be described as a Hamiltonian system on a Poisson
manifold. Therefore, we are in a situation similar to a standard
classical system.  We have seen that the dynamics preserves the
submanifold $M_C\times S_Q$ ($S_{Q}$ being the sphere  in Eq. (\ref{eq:32})),
and thus it is natural to consider such a manifold as the space of our
statistical system. 

We have thus to construct now a statistical system composed of two
subsystems. We may think in introducing a total distribution 
factorizing as the product of a classical and a quantum distribution. But as the
events of the classical and the quantum regimes are not independent,
the two probabilities must be combined and not defined through 
two factorizing functions.

Hence we shall consider a density $F_{QC}\in
\mathcal{O}$ and the canonical volume element $d\mu_{QC}$ in the
quantum-classical phase space, restricted to $M_C\times S_Q$:
\begin{equation}
\label{eq:volumeform}
d\mu_{QC}=d\mu_Cd\Omega_Q,
\end{equation}
where $d\mu_C$ is given by (\ref{eq:volumeclassical})
and $d\Omega_Q$ is the volume element on a $(2N_Q-1)$-dimensional sphere
(see below).

With these ingredients we introduce the following 
\begin{definition}
  Given an observable $M\in \mathcal{O}$, we define the macroscopic
  average value, $\langle M\rangle$, as
  \begin{equation}
    \label{eq:15}
    \langle M\rangle=\int_{M_C\times S_Q}\hspace{-0.2cm}M(\vec R, \vec P, \vec q, \vec
    p) F_{QC}(\vec R, \vec P, \vec q, \vec p)d\mu_{QC}.
  \end{equation}
\end{definition}
In the rest of the section we shall discuss in more detail the ingredients
of this definition. 

We start by the volume form. Given the classical
phase space $M_C$ with Darboux coordinates $(\vec R,\vec P)$
(they always exist locally\cite{Abraham:1978p1131}) we define the volume 
element in the classical phase space by
\begin{equation}
\label{eq:volumeclassical}
d \mu_C=\prod_JdP_JdR_J,
\end{equation}
that, as it is well known, is invariant under any purely classical
Hamiltonian evolution.
One may wonder  to what extent the volume form depends on the choice of 
coordinates. But taking into account that two different sets of Darboux 
coordinates differ by a canonical transformation, and using the fact that these
transformations always have Jacobian equal to one, one concludes 
that the volume form does not depend on the choice of coordinates.

We might proceed in the same way for the quantum part of our system,
as we have written the quantum dynamics like a classical Hamiltonian one
and, therefore, we could use the canonical invariant
volume element describe in the previous paragraph $d\mu_Q$ on $M_Q$.
However, new complications appear in this case as in the quantum part
of the phase space we restrict the integration to $S_Q$ and then we look 
for an invariant volume form in the unit sphere, rather than in the full 
Hilbert space. 

The new ingredient that makes the restriction possible is the fact
that all our observables and of course, also the Hamiltonian,
are killed by $\Gamma$. This implies, 
as discussed before, that $f_{\mathbb I}$ is
a constant of motion.
Using this we decompose
$$
d\mu_Q= df_{\mathbb I}d\widetilde\Omega_Q,
$$
which is possible on ${\cal H}\setminus\{0\}$. 
The decomposition is not unique, but the restriction (pullback) 
of  $d\widetilde\Omega_Q$ to the unit sphere gives a uniquely defined, 
invariant volume form on $S_Q$ that we will denote by $d\Omega_Q$.
The decomposition is nothing but the factorization into the radial 
part and the solid angle volume element.

\begin{example}
  For the simple case we studied above, where
$\mathcal{H}=\mathbb{C}^2$ and hence $M_Q=\mathbb{R}^4$ and $S_Q$ is a
three dimensional sphere, the volume element above would be:
\begin{align}
    \label{eq:28b}
    d\Omega_{Q}=\;&q_1dq_2dp_1dp_2-q_2dq_1dp_1dp_2 
    \notag
    \\
    +&p_1dq_1dq_2dp_2-p_2dq_1dq_2dp_1.
  \end{align}
\end{example}

Now, we finally put together the two ingredients to obtain
the invariant volume element $d\mu_{QC}$ in eq. (\ref{eq:volumeform}).

We discuss now the properties we must require from $F_{QC}$ in order
for  Eq. (\ref{eq:15}) to correcly define the statistical mechanics for the
Ehrenfest dynamics.

\begin{itemize}
  \item The expected value of the constant observable should be that constant, 
        which implies that the integral on the whole set of states is equal to one:
        \begin{equation}
         \label{eq:16}
         \int_{M_C\times S_Q}F_{QC}d\mu_{QC}=1.
        \end{equation}
  \item The average, for any observable $f_A$ of the form (\ref{eq:42}) 
        associated to a positive definite hermitian operator $A$ should 
        be positive. This implies the usual requirements of positive 
        probability density in purely classical statistical mechanics.
\end{itemize}

So far we have discussed general considerations about the statistical 
mechanics of our system. Now we should include the dynamics in the game.
Given that we have formulated the Ehrenfest dynamics as a Hamiltonian
system, it is well known \cite{balescu1997statistical} which is the evolution 
of the density function $F_{QC}$. It is given by the {\bf Liouville equation}
\begin{equation}
  \label{eq:19}
  \frac{dF_{QC}}{dt}+\{f_H, F_{QC}\}=0,
\end{equation}
where, in the derivation of the equation, it is an essential requirement
the invariance of the volume form under the evolution of 
the system.

The equilibrium statistical mechanics is obtained by requiring $\dot F_{QC}=0$, 
which making use of (\ref{eq:19}) is equivalent to
$$\{f_H, F_{QC}\}=0,$$
or, in other words, $F_{QC}$ should be a constant of motion.
An obvious non trivial constant of motion is
given by any function of $f_H$. The question is 
if there are more.

To answer this we should examine closely our particular 
dynamics. In several occasions we have mentioned the non linear 
character of a generic Ehrenfest system. This is an essential
ingredient, because due to this fact we do not expect any other 
constant of motion that  a function of the Hamiltonian itself.
This is in contrast with the case of linear equations of motion,
where the system is  necessarily integrable.

The question of the existence of more constants of motion and, therefore,
more candidates for the equilibrium density $F_{QC}$ can be elucidated with a simple example that we discuss below. 
In it we show the presence of ergodic regions (open regions in the leave of constant energy densely covered by a single orbit). 
This rules out the existence of constants
of motion different from the constant function in the region
(i.e., functions of $f_H$).

\begin{example}
In the following example we will study a simple toy model
in which the coupling between classical and quantum degrees of freedom
gives rise to chaotic behavior and the appearence of ergodic regions.

The system consists of a complex two dimensional Hilbert space
$M_Q=\mathbb{C}^2$ and a classical 2-D phase space where we define a 1-D harmonic 
oscillator. Using coordinates $(I_{\theta},\theta)$ for the classical variables (action-angle 
coordinates for the oscillator)
and $\Psi\in\mathbb{C}^2$ we define the following Hamiltonian
$$f_H=I_{\theta} + {\frac 1 2}\langle
\Psi\vert\sigma_z
+\epsilon \cos(\theta)\sigma_x\vert\Psi\rangle,$$
with $\sigma_x, \sigma_z$ the Pauli sigma matrices. 

We parametrize the normalized quantum state by
$$|\Psi\rangle=e^{i\alpha}
\begin{pmatrix}I_{\phi} \\ 
e^{i\phi}\sqrt{1-I_{\phi}^2}
\end{pmatrix}.$$
One can now easily check that $I_{\phi}$ and $\phi$ are canonical conjugate
variables.

In these variables the Hamiltonian reads
\begin{equation}
  \label{eq:44}
  f_H= I_{\theta}+ I_{\phi}^2+{\epsilon}I_{\phi}\sqrt{1-I_{\phi}^2}
\cos(\theta)\cos(\phi),
\end{equation}
where $\epsilon$ measures the coupling of the classical and
quantum systems.

In the limit of vanishing $\epsilon$ the system
is integrable and actually linear in these coordinates.
However for non vanishing $\epsilon$ the model becomes non linear
and, as we will show below, regions of chaotic motion emerge.

In order to understand the behaviour of our system it is useful
to study its Poincar\'e map (see 
Ref. \onlinecite{Abraham:1978p1131} for the definition). To this end we take the 
transversal (or Poincar\'e)  section at 
$\theta=0$ and, taking into account the conservation of energy,
only two coordinates are needed to describe the map, we have chosen
the quantum variables $I_{\phi}$ and $\phi$.

\begin{figure}[t!]
  \centering
\includegraphics[width=7cm]{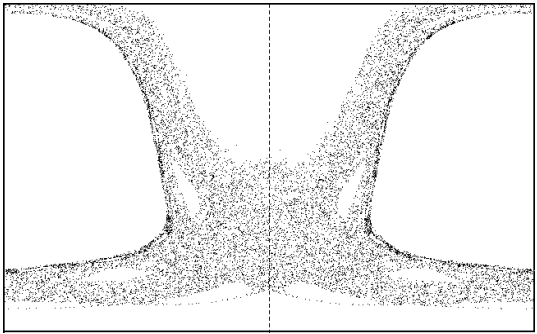}

  \caption{The plot shows a single orbit of the Poincar\'e map   of
Eq. \eqref{eq:44} at $\theta=0$.  The 
angle $\phi\in[-\pi,\pi]$ is plotted in the horizontal axis and 
$I_{\phi}\in[0,1]$ in the vertical one.
We have taken $\epsilon=0.8$. The energy determines $I_{\theta}$  and its actual
value does not affect the dynamics of $\phi$ and $I_{\phi}$.}
\label{fig:1}
\end{figure}

By examining the plot of Fig. \ref{fig:1}, one sees that the orbit
densely fills a large region of the 
Poincar\'e section and therefore any constant of motion should be the
constant function in that region.
In the presence of such an ergodic evolution the only constants of motion
are functions of the Hamiltonian.
\end{example}

As a consequence we can claim that, genericaly within the Ehrenfest dynamics,
the only functions which commute with
the Hamiltonian in \eqref{eq:19} are those which are function of the
Hamiltonian $f_H$ itself. As the Poisson bracket is skew-symmetric,
the property follows trivially. Ergodicity helps us to assert, roughly
speaking, that in general no other function in $\mathcal{O}$ will be a constant
of the motion.

Thus, at this stage, we can finish our construction with the
equilibrium distribution associated to Ehrenfest dynamics. Taking into
account the ergodicity of the dynamics, we can impose the
equal-probabilities condition to the configurations of our isolated system, which leads to 
the microcanonical ensemble, (see Ref. \onlinecite{Oliveira2007ergodic}). 
Then, the probability density for the canonical ensemble 
turns out to be:
\begin{equation}
  \label{eq:31}
  F_{QC}^{c}(\vec R, \vec P, \vec q, \vec p)=Z^{-1}e^{-\beta f_H},
\end{equation}
where $Z=\int_{M_C\times S_Q}e^{-\beta f_H}d\mu_{QC}$, 
as it can be seen, in a standard derivation, for example in Refs. \onlinecite{balescu1997statistical,balescu1975equilibrium, Schwabl2002statistical}, where $\beta^{-1}$, being proportional to the temperature of the 
isolated system, is the parameter governing the equilibrium between the different parts of 
the system. An
information-theoretic approach to the equiprobability in 
the microcanonical ensemble and to Eq. (\ref{eq:31})  may be seen in Ref. \onlinecite{Jaynes1957maxent1} or
in Refs. \onlinecite{Ellis1985largedev,Touchette2009largedev} for a much more modern and mathematically sound
exposition.

%\fbox{To finish}: Implications, relation with the trace of the quantum
%system if we consider the subdistribution $\tilde \rho$ ...
 
\section{Conclusions and future work}
\label{sec:concl-future-work}
In this paper we have constructed a rigorous Hamiltonian description
of the Ehrenfest dynamics of an isolated system by combining the Poisson 
brackets formulation of classical mechanics with the geometric formulation of quantum
mechanics. We have also constructed the corresponding statistical
description and obtained the associated Liouville equation.  
Finally, after verifying numerically that the Ehrenfest dynamics is ergodic, we
justify the equilibrium distribution produced by it.

The definition of the Hamiltonian in Eq. \eqref{eq:9} and its associated
canonical equilibrium distribution \eqref{eq:31} allows us now to use (classical) Monte
Carlo methods for computing canonical equilibrium averages in our
system, given by the expression \eqref{eq:15}. 

Also it is straightforward to extend the non-stochastic method
proposed by Nos\'e \cite{nose:1984,nose:1991} to our Hamiltonian \eqref{eq:9}. 

However, if we wish to perform MD simulations using stochastic
methods, we should first construct the
Langevin dynamics associated with our Hamiltonian \eqref{eq:9} and equations
\eqref{eq:6}-\eqref{eq:6-6}. To develop this program, we
also need an analogue of the Fokker-Planck equation associated with those
Langevin equations and then we have to check whether its solution at infinite time
approaches \eqref{eq:31}. Presently we are working on this
point \cite{futurepaper}, 
i.e., on the extension of our formalism to an associated stochastic MD
that can be used to rigorously probe non-equilibrium phenomena at
constant temperature.

\begin{acknowledgments}
 We would like to thank Jos\'e F. Cari\~nena, Andr\'es Cruz and David Zueco for many illuminating 
discussions.

This work has been supported by the research projects E24/1 and E24/3 (DGA,
Spain), FIS2009-13364-C02-01 (MICINN, Spain) and 200980I064 (CSIC,
Spain). 
\end{acknowledgments}

%\bibliography{chemistry}
%\bibliographystyle{}
%

\end{document}